\documentclass[aps,prb,preprint,psfig,groupedaddress]{revtex4-1}

\usepackage{amsmath,amssymb}
\usepackage{graphicx}
\usepackage{bm}
\usepackage{epstopdf} 

\AppendGraphicsExtensions{.tif}

\newcommand{\BS}{Bi$_2$Se$_3$}
\newcommand{\BST}{(Bi,Sb)$_2$Te$_3$}
\newcommand{\AlO}{Al$_2$O$_3$}
\newcommand{\STO}{SrTiO$_3$}
\newcommand{\SiN}{Si$_3$N$_4$}

\renewcommand{\vec}[1]{\mbox{\boldmath$1$}}

\newcounter{lastnote}

\def\bc{\begin{center}}
\def\ec{\end{center}}
\def\be{\begin{equation}}
\def\ee{\end{equation}}
\renewcommand{\vec}[1]{\mbox{\boldmath$1$}}

\begin{document}
\title{Mapping the chemical potential dependence of current-induced spin polarization in a topological insulator}
\author{Joon Sue Lee$^1$, Anthony Richardella$^1$, Danielle Reifsnyder Hickey$^2$, K. Andre Mkhoyan$^2$, and Nitin Samarth$^1$}
\email{nsamarth@psu.edu}
\affiliation{$^1$Department of Physics, The Pennsylvania State University, University Park, Pennsylvania 16802, USA}
\affiliation{$^2$Department of Chemical Engineering and Materials Science, University of Minnesota, Minneapolis, Minnesota 55455, USA}

\date{\today}
\begin{abstract}
We report electrical measurements of the current-induced spin polarization of the surface current in topological insulator devices where contributions from bulk and surface conduction can be disentangled by electrical gating. The devices use a ferromagnetic tunnel junction (permalloy/Al$_2$O$_3$) as a spin detector on a back-gated \BST{} channel. We observe hysteretic voltage signals as the magnetization of the detector ferromagnet is switched parallel or anti-parallel to the spin polarization of the surface current. The amplitude of the detected voltage change is linearly proportional to the applied DC bias current in the \BST{} channel. As the chemical potential is tuned from the bulk bands into the surface state band, we observe an enhancement of the spin-dependent voltages up to 300\% within the range of the electrostatic gating. Using a simple model, we extract the spin polarization near charge neutrality (i.e. the Dirac point).
\end{abstract}
\maketitle

\section{Introduction}

The combination of strong spin-orbit coupling, time-reversal symmetry and inversion symmetry is known to create gapless helical, Dirac surface states that lie within the bulk band gap of narrow band gap semiconductors such as the Bi-chalcogenides. These three-dimensional (3D) ``topological insulators" \cite{hasan2010,qi2010,moore2010} have begun to attract significant contemporary interest for spintronics since the spin-momentum locking in the surface state endows an inherent spin polarization to surface state charge currents. \cite{Pesin2012,Mahfouzi2012} This has motivated several recent experiments on ``topological spintronics." Spin-transfer torque has been demonstrated in bilayers of ferromagnets and topological insulators, \cite{Mellnik2014,Fan2014,Wang2015} with record values of the spin torque ratio at room temperature in \BS\cite{Mellnik2014}and at low temperature in \BST . \cite{Fan2014}  Spin Hall angles have also been measured through the inverse spin Hall effect in topological insulators, created by the injection of a spin current from a metallic ferromagnet through spin pumping \cite{Jamali2014,Deorani2014,Shiomi2014}and through spin-polarized tunneling. \cite{Liu2015}

The spin polarization of the surface states of topological insulators has been measured in extensive spin- and angle-resolved photoemission spectroscopy studies.\cite{Hsieh2009,Xu2011,Jozwiak2013,Neupane2014} However, direct electrical measurements, relevant for developing spintronic devices, are just beginning to emerge and derive from a well-established methodology in spintronics: the essential idea is to use a ferromagnetic contact as a voltage probe of the spin-dependent electrochemical potential.  As the relative orientation of the current spin polarization and ferromagnet magnetization changes, so does the measured voltage. \cite{Hong2012} This scheme was first used to clearly demonstrate spin-momentum locking using ferromagnetic tunnel contacts on \BS{} transport channels, albeit with the chemical potential of \BS{} in the bulk conduction band and at temperatures below 150 K. \cite{Li2014} More recent experiments have extended such measurements to room temperature in \BS{}\cite{Dankert2014} and to topological insulators with reduced bulk conduction.\cite{Tang2014,Ando2014,Tian2015} Further, a recent study used spin-polarized tunneling in ferromagnet/insulator/topological insulator junctions to map out the energy-dependence of the effective spin polarization in \BS.\cite{Liu2015} Here, we demonstrate an experiment that is complementary to these prior studies. We show that the standard three probe spin potentiometry scheme using a magnetic tunnel junction (MTJ)\cite{Hong2012,Li2014} can be effectively applied to electrically gated topological insulator devices, thus yielding insights into the chemical potential dependence of the spin polarization of the surface state current. In our measurements, we map out the spin-dependent voltage as the chemical potential is tuned from the bulk conduction band through the Dirac point and into the valence bulk band. We find that the spin-dependent voltages are enhanced as the chemical potential is moved away from the bulk conduction/valence bands into the surface state band. We also show that we do not observe possible contributions of the opposite spin polarization from Rashba spin-split bands within the range of the electrostatic gating.

\section{Sample synthesis and device fabrication}

\BST{} thin films were grown on \STO{} substrates by molecular beam epitaxy (MBE).\cite{Richardella2015} We used a Bi:Sb ratio of $\sim$1:1 to place the chemical potential into the bulk band gap. The samples had film thicknesses of 7 nm (S1), 8 nm (S2), and 6 nm (S3); this minimizes contributions from unpolarized bulk carrier conduction and allows the tuning of the electron chemical potential of both the bottom and top surfaces by back gating. We note that the film thicknesses are above the limit where Dirac point is disrupted by hybridization of the top and bottom surface states\cite{Neupane2014,Li2010,Jiang2012}. 

\begin{figure}
\includegraphics[width=85mm]{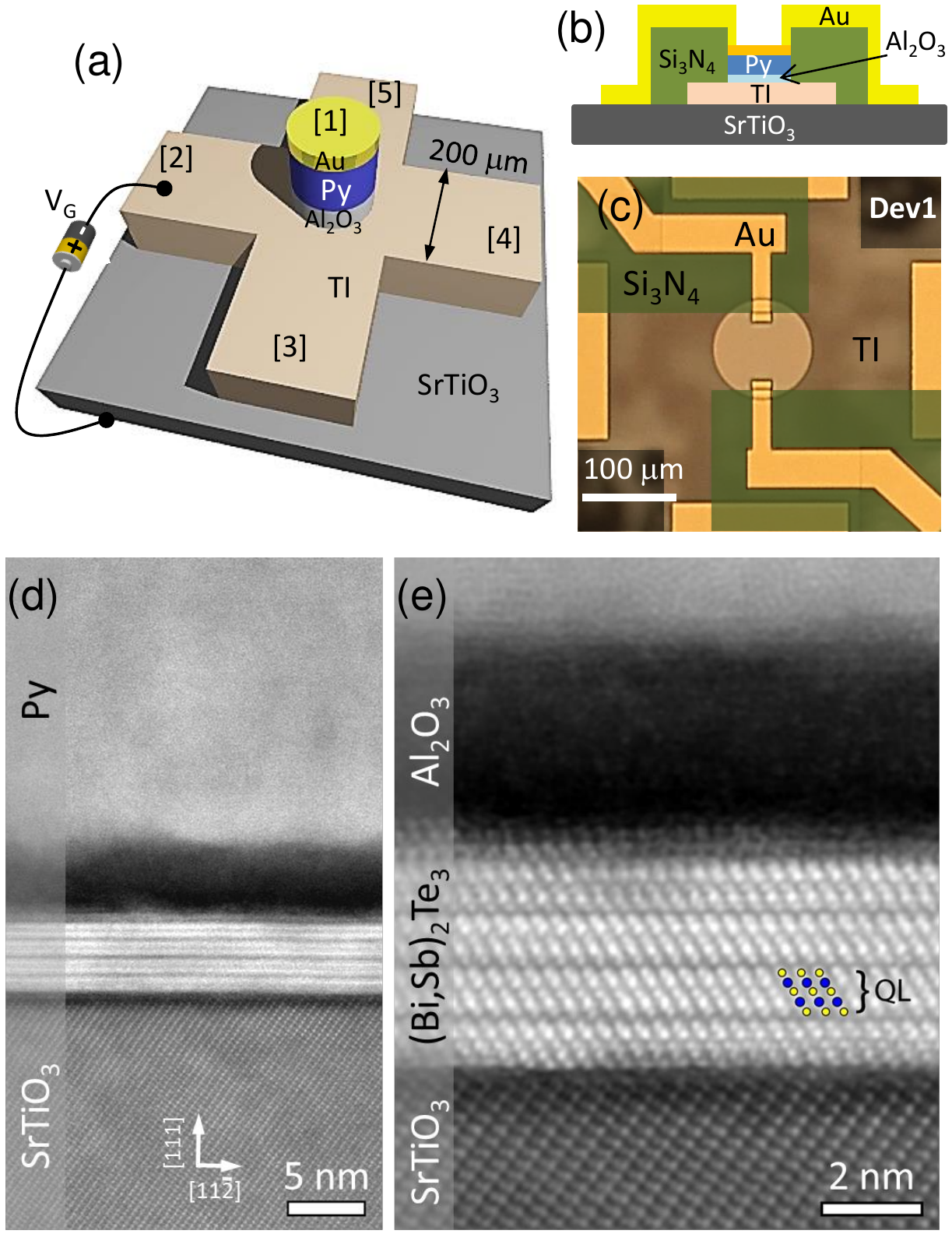} 
\caption{(Color online) (a) Device schematic of an \AlO{}/Py MTJ on a cross-shaped topological insulator channel. Contact leads are labeled as 1 through 5. (b) A cross-section of an MTJ is illustrated. (c) An optical microscope image of a device with a circular-shaped MTJ of 50 $\mu$m radius (Dev1). \SiN{} layer is false-colored in green. (d) A cross-sectional HAADF-STEM image of the \AlO{}/Py MTJ on \BST{} (Dev5), viewed along the [1 $\bar{1}$ 0] direction of the \STO{} substrate. (e) Atomic-resolution HAADF-STEM image of the \BST{} structure, also viewed along the [1 $\bar{1}$ 0] direction of the substrate. STEM images were FFT low-pass filtered to 1.5 \AA (0.67 \AA$^{-1}$ in reciprocal space) resolution to remove instrumental noise. (Bi,Sb) atoms are indicated in blue, Te atoms are indicated in yellow, and 1 quintuple layer (QL) is labeled.}
\label{fig1}
\end{figure}

For clean measurements of the surface state spin polarization, we developed a device geometry that restricts all measurements to the top surface [Fig.\ref{fig1}(a)]. We first define a cross-shaped \BST{} channel by standard photolithography followed by Ar plasma etching. A bilayer photoresist window of circular or rectangular shape is then patterned at the center of the cross-shaped channel. The bare surface of the topological insulator is cleaned using a low-power Ar plasma etch (15 W for 20 s) that removes possible native oxides and photoresist residues along with $\sim$1 nm \BST{}. Following the sputter deposition of an \AlO{} seed layer ($\sim$0.3 nm), we use atomic layer deposition (ALD) for conformal deposition of an \AlO{} layer (2.1 nm for devices on S1 and 1.8 nm for devices on S2) over the topological insulator surface. A Py layer (20 nm) and a thin Au capping layer (5 nm) are then deposited by e-beam evaporation, followed by a lift-off process. Finally, the top of the MTJ is extended to two Au contact pads over a 60-nm-thick \SiN{} layer to isolate the MTJ from the topological insulator channel as shown in Figs.\ref{fig1}(b) and \ref{fig1}(c). Each sample has multiple devices with various MTJ areas. The five devices studied in this paper are listed in Table I.

For microstructural analysis, a device was cross-sectioned using a focused ion beam (FIB, FEI Quanta 200 3D) and then imaged using high-angle annular dark-field scanning transmission electron microscopy (HAADF-STEM) [Figs.\ref{fig1}(d) and \ref{fig1}(e)]. For STEM imaging, an aberration-corrected (CEOS DCOR probe corrector) FEI Titan G2 60-300 S/TEM equipped with a Schottky X-FEG gun was operated at 200 kV with a probe convergence angle of 16 mrad. Additionally, the \AlO{} layer was identified as being amorphous by conventional TEM imaging (not shown), using an FEI Tecnai G2 F30 at 300 kV. 

\section{Results and discussions}

\subsection{Ambipolar transport of \BST{}}

\begin{figure}
\includegraphics[width=80mm]{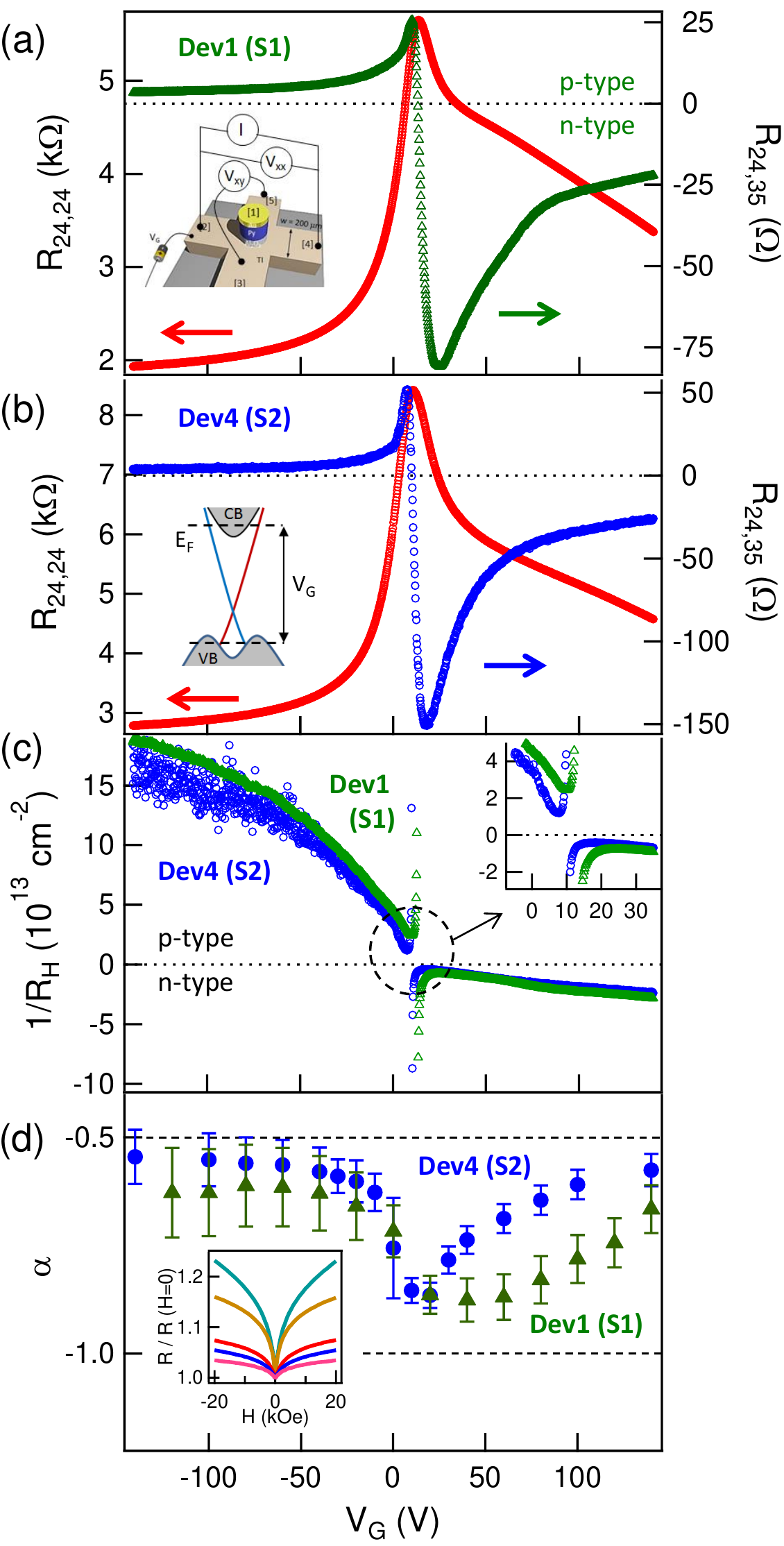} 
\caption{(Color online) (a) Gate voltage dependence of the two-terminal channel resistance (red circles) at zero field and Hall resistance (green triangles) at $H = 10$ kOe from Dev1 (S1). Inset: measurement setup. (b) Similar measurements as (a) for Dev4 (S2). Inset: schematic of the band structure of \BST{} in presence of $V_G$. (c) Inverse of the Hall resistance for Dev1 (green triangles) and for Dev4 (blue circles) in units of 2D carrier density (cm$^{-2}$). Inset: zoom-in near the charge neutrality point ($V_G=13$ V for Dev1 and $V_G=10$ V for Dev4). (d) Variation of prefactor $\alpha$ with $V_G$, obtained by fitting magneto-resistance (inset) to Eq. (1) for Dev1 (green triangles) and for Dev4 (blue circles). Error bars represent an uncertainty with 95 percent confidence in data fitting. Inset: magneto-resistance for Dev4 at $V_G= -140$ V (pink), $-20$ V (blue), 20 V (cyan), 60 V (orange), 140 V (red). All data from Dev1 (Dev4) were taken at 2.1 K (1.8 K).}
\label{fig2}
\end{figure}

We first discuss the back-gate-voltage ($V_G$) dependence of the channel resistance and the Hall resistance. The former is measured through the two-terminal channel voltage between leads 2 and 4 and the latter between 3 and 5 is measured while a DC current of 1 $\mu$A flows between leads 2 and 4 [see Fig.\ref{fig1}(a) for lead configuration].  The channel between two Ti/Au leads (between 2 and 4 as well as between 3 and 5) is 300 $\mu$m long by 200 $\mu$m wide, and the contact resistance of each lead is $\sim$100 $\Omega$. The Hall voltage was measured as a function of applied perpendicular-to-plane magnetic field of 10 kOe. The results show the typical signatures of ambipolar transport in a topological insulator: the change of carrier type from electrons to holes as well as a peak in channel resistance near the charge neutrality point (at $V_G=20$ V for S1 and at $V_G=10$ V for S2) [Figs.\ref{fig2}(a) and \ref{fig2}(b)]. Carrier densities at each end of the applied gate-voltage range were obtained to be $n_e = 3.53\times 10^{13}$ ($2.44 \times 10^{13}$) cm$^{-2}$ at $V_G = 140$ V and $n_p = 1.91\times 10^{14}$ ($1.69\times 10^{14}$) cm$^{-2}$ at $V_G=-140$ V for S1 (S2), respectively [Fig.\ref{fig2}(c)]. These values give an idea of the range of the chemical potential tuning. We believe that we were able to change the chemical potential across the bulk band gap from a position near the bottom of the conduction band (at $140$ V) to a position below the top of the valence band (at $-140$ V). 

\subsection{Weak anti-localization of \BST{}}

To further understand the surface and bulk conduction in the \BST{} films with respect to the gate voltage, we carried out magneto-transport measurements using the same wiring configuration. Near 2 K, the magneto-resistance of a topological insulator channel shows a sharp cusp near zero magnetic field as shown in the inset to the Fig.\ref{fig2}(d). This suppression of resistance near zero magnetic field can be explained by weak anti-localization, typically seen in conventional 2D metals with strong spin-orbit coupling\cite{Suzuura2002} as well as in 3D topological insulators that have a non-trivial $\pi$ Berry's phase\cite{Hsieh2009}. The quantum corrections to the diffusive magneto-conductance are given by\cite{Hikami1980}
\begin{equation}
\sigma(H)-\sigma(0)=\alpha\frac{e^2}{2\pi^2\hbar}\left[\psi\left(\frac{1}{2}+\frac{\hbar c}{4el^2_\phi H}\right)-\ln \left(\frac{\hbar c}{4el^2_\phi H}\right)\right].
\label{HLN}
\end{equation}
Here, $\sigma$ is the 2D channel conductivity and can be determined from the 2D channel resistivity $\rho_{xx}$ and 2D Hall resistivity $\rho_{xy}$ via $\sigma = \rho_{xx}/(\rho_{xx}^2+\rho_{xy}^2)$. Also, $e$, $\hbar$, $\psi (x)$, $\alpha$, and $l_{\phi}$ are the electronic charge, the Planck's constant, the digamma function, the pre-factor, and the coherence length, respectively. Different types of localization behaviors can be identified by the values of the pre-factor $\alpha$, which takes on values of $1$, $0$, $-1/2$ for the orthogonal, unitary and symplectic classes, respectively.\cite{Hikami1980} The weak anti-localization behavior of the surface state of a 3D topological insulator is in the symplectic class with $\alpha = -1/2$ expected for one channel conduction of a surface state or bulk-carrier-coupled top and bottom surfaces. For two conduction channels of decoupled top and bottom surfaces, $\alpha =(-1/2)+(-1/2)=-1$\cite{Garate2012,Checkelsky2011,Steinberg2011}. 

Figure \ref{fig2}(d) shows the gate-voltage dependence of the pre-factor $\alpha$ obtained from the fitting of weak anti-localization by Eq.(1) at different gate voltages. At each end of the applied gate-voltage range ($-140$ V and $140$ V), $\alpha$ approaches $-1/2$, indicating that the chemical potential is on or near the bulk bands whose carriers couple the top and bottom surfaces. Near the charge neutrality point where the bulk conduction is minimal, $\alpha$ deviates from $-1/2$ and approaches $-1$, indicating that the surface states on the top and bottom surfaces have become decoupled, independent conduction channels. $\alpha$ changes faster with the chemical potential as it is lowered from the charge neutrality point towards the valence band because the Dirac point of \BST{} is closer to the top of the valence band than the bottom of the conduction band. We note that the contact resistance was subtracted before fitting the two-terminal magneto-resistance data to Eq.(1).

\subsection{Tunnel junction properties}

\begin{figure*}
\includegraphics[width=165mm]{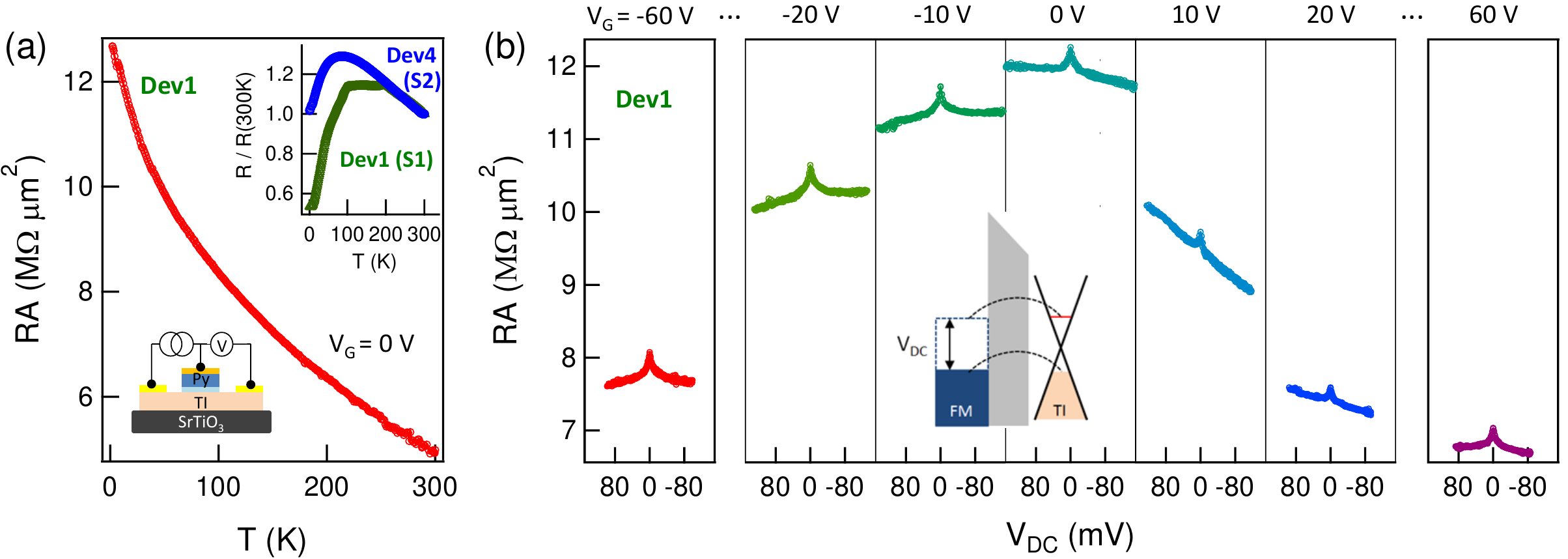} 
\caption{(Color online) (a) Temperature dependence of the RA product shows an insulating behavior. Bottom inset illustrates a schematic of three-terminal measurement setup. Top inset shows temperature dependence of the channel resistance of Dev1 (S1) and Dev4 (S2). (b)$V_{DC}$ dependence of RA product from Dev1 at 1.8 K with different gate voltages. Middle inset illustrates the tunneling across the \AlO{} tunnel barrier between $V_{DC}$-tuned ferromagnetic metal and topological insulator top surface.}
\label{fig3}
\end{figure*}

The RA product of the tunnel junctions is in the range 10$^6$-10$^7$ M$\Omega\cdot\mu$m$^2$. We observe qualitatively similar behavior of the gate-voltage dependence, the tunnel junction characteristics, and the spin-dependent voltages from all the devices. Typically, the temperature dependence of the zero-bias junction resistance shows an insulating behavior [Fig.\ref{fig3}(a)], while the channel resistance of the topological insulator shows metallic behavior below 100 K [Fig.\ref{fig3}(a) inset]. The junction resistance as a function of measured DC voltage ($V_{DC}$) across an MTJ shows an anomaly at zero-bias [Fig.\ref{fig3}(b)], commonly seen in MTJs\cite{Akerman2002}.
We also observed an asymmetry in the junction resistance between positive and negative voltages [Fig.\ref{fig3}(b)]. Interestingly, the asymmetric behavior qualitatively follows the gate-voltage dependence of the channel resistance [Fig.\ref{fig2}(a)], which can be interpreted as probing of the density of states of the topological insulator top surface via tunneling between the $V_{\rm{DC}}$-tuned metal and the $V_G$-tuned topological insulator.

\subsection{Spin-dependent voltages from topological insulator spin polarization}

\begin{figure*}
\includegraphics[width=140mm]{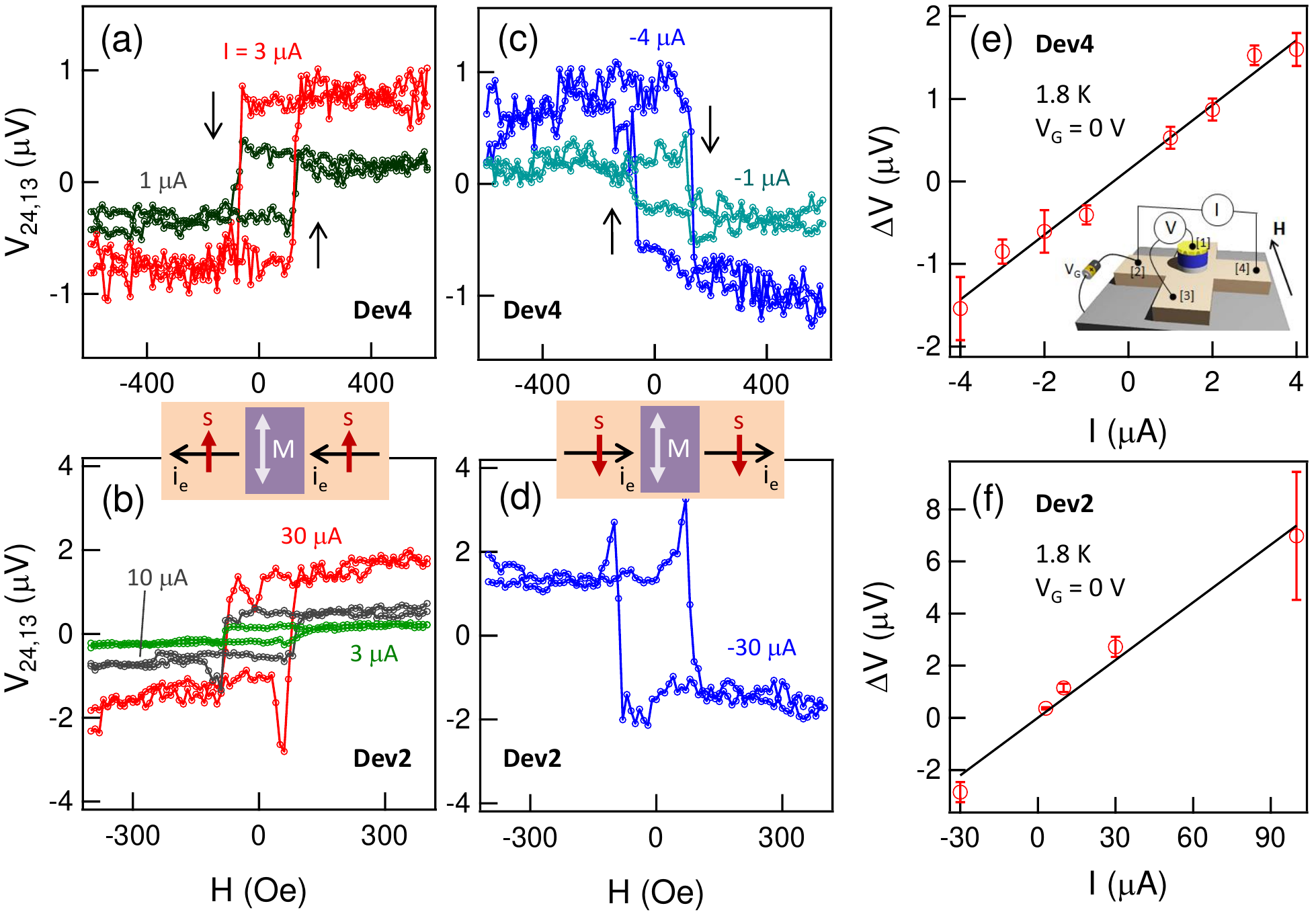} 
\caption{(Color online) Detected spin-dependent voltages with positive DC bias current (a) from Dev4 [3 $\mu$V (red) and 1 $\mu$V (green)] and (b) from Dev2 [30 $\mu$V (red), 10 $\mu$V (gray), and 3 $\mu$V (green)]. Reversed spin-dependent voltages with negative bias currents (c) from Dev4 [$-4$ $\mu$V (blue) and $-1$ $\mu$V (cyan)] and (d) from Dev2 [$-30$ $\mu$V (blue)]. The schematics depict spin polarization (red arrow) induced by electron current (black arrow) on a surface of a topological insulator channel and the magnetization of the ferromagnetic layer (white arrow). The voltage change $\Delta V$ as a function of bias current (e) from Dev4 and (f) from Dev2 shows a linear relationship. The black line is a linear fit to data shown as red circles with error bars that represent the standard deviations over multiple measurements. Inset in (e) illustrates the measurement setup. All data were taken at 1.8 K with $V_G$= 0 V.}
\label{fig4}
\end{figure*}

Our three-terminal potentiometric measurement scheme follows the proposal by Hong {\it et al.}\cite{Hong2012} We use an \AlO{}/Py junction for the ferromagnetic electrode while a dc current flows through a topological insulator channel with an in-plane magnetic field perpendicular to the current direction. When the direction of the Py magnetization switches from parallel to anti-parallel to the current-induced spin polarization of the topological insulator surface state, we observe step-like changes in the detected voltage, resulting in hysteretic spin-dependent voltages associated with the coercive field ($\sim$100 Oe) of the Py layer. For positive currents as shown in Figs.\ref{fig4}(a) and \ref{fig4}(b), the relative direction between the Py magnetization and the topological insulator spin polarization determines the positive change in the detected voltage $\Delta V= [V({\bf M})-V(-{\bf M})]$. For reversed currents [Figs.\ref{fig4}(c) and \ref{fig4}(d)], the direction of topological insulator spin polarization is opposite to the case of positive currents, so that the change in the detected voltage $\Delta V$ becomes negative. 
Figures \ref{fig4}(e) and \ref{fig4}(f) show that the measured voltage change is linearly proportional to the bias currents. This linear relation can be expressed by
\begin{equation}
\Delta V = I R_B (\bf{p} \cdot \bf{m}),
\label{Vchange}
\end{equation}
where $R_B$, $\bf{p}$, and $\bf{m}$ represent ballistic resistance, degree of the spin polarization per unit current in the topological insulator channel, and effective magnetic polarization of ferromagnet, respectively.\cite{Hong2012} Depending on the position of the chemical potential tuned by the electrical gating, the measured spin-dependent voltages may originate not only in the topological insulator surface states but also in the Rashba spin-split bands. Equation (2) is applicable to the spin-dependent voltages from combined conduction channels including the surface states, the Rashba spin-split bands, and the unpolarized bulk bands in both ballistic and diffusive limits.
 
\subsection{Gate-voltage dependence of spin-dependent voltages}

\begin{figure*}
\includegraphics[width=175mm]{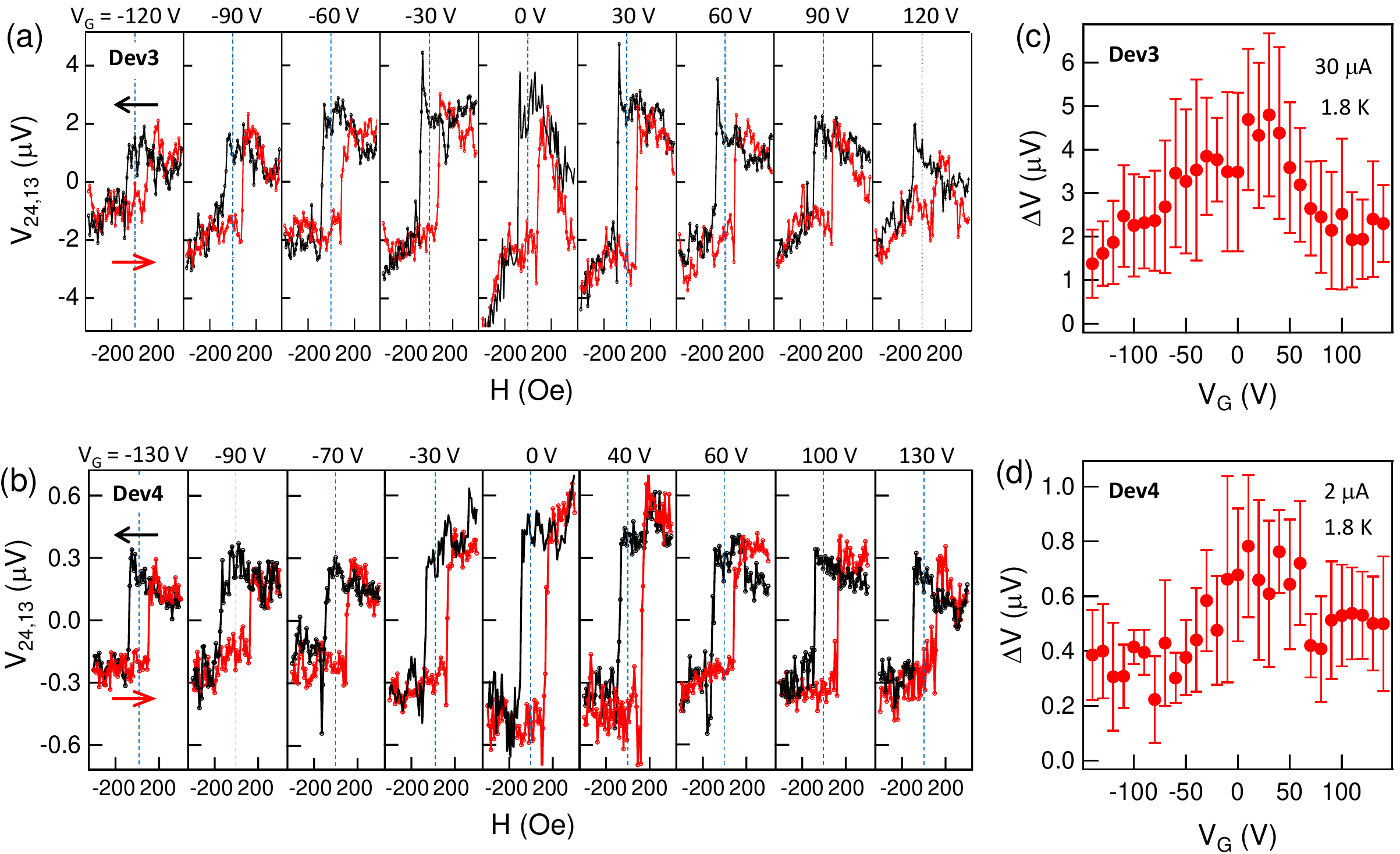} 
\caption{(Color online) Detected hysteretic spin-dependent voltages with up sweep (red) and down sweep (black) of in-plane magnetic field with different gate voltages (a) from Dev3 and (b) from Dev4. Observed $\Delta V$ with respect to $V_G$ is plotted in (c) for Dev3 and in (d) for Dev4. The error bars represent the standard deviations over multiple measurements. All data were taken with fixed currents of $30$ $\mu$A for Dev3 and $2$ $\mu$A for Dev4 at 1.8 K.}
\label{fig5}
\end{figure*}

We now measure the hysteretic spin-dependent voltages while changing the chemical potential across the bulk bandgap. When the chemical potential is placed in the bulk band gap, the observed spin-dependent voltages are two to three times larger. Using two devices (Dev3 from S1 and Dev4 from S2), multiple measurements of magnetic field sweeps, repeated at least 8 times, were carried out every 10 V from 140 V to $-140$ V. Figures \ref{fig5}(a) and \ref{fig5}(b) show the hysteretic spin-dependent voltages at some of the $V_G$'s among the multiple measurements. As shown in Figs.\ref{fig5}(c) and \ref{fig5}(d), the magnitude of the resulting voltage change $\Delta V$ is maximal near the charge neutrality point ($V_G=$13 V for S1 and $V_G=$10 V for S2).  

We do not observe any sign change in $\Delta V$ throughout the $V_G$ range, suggesting that with a fixed current bias, the carrier type change from n to p does not flip the direction of the measured spin polarization. This observation is fully consistent with dominant contributions to the spin polarization originating from the helical Dirac surface states since we use a dc current for the measurement. For a given crystal momentum, the spin polarization of holes is opposite to that of electrons. However, with a fixed DC current, the propagation direction of holes (below the Dirac point) is the reverse of the propagation of electrons (above the Dirac point). Therefore, the opposite spin polarization and the opposite propagation direction between electrons and holes make the sign of the detected spin-dependent voltages remain the same above and below the Dirac point.\cite{Liu2015}  We also note that our observations do not reveal any significant contributions from co-existing Rashba states which would have resulted in the opposite behavior due to the spin polarization being opposite to that of topological surface states \cite{Hong2012}. For the latter, we assume that the Rashba bands minimum is close to the bottom of the conduction band and the chemical potential is tuned above the Rashba bands minimum. It is also possible that in the measurement range the chemical potential does not lie on any Rashba spin-split bands. 

\subsection{Spin polarization calculation}

Equation (2) can be re-written as $\Delta V = I R_B P_{\rm{TI}} P_{\rm{FM}}$, where $P_{\rm{TI}}$ is the spin polarization of the topological insulator and $P_{\rm{FM}}$  is the magnetic polarization of the ferromagnetic detector. The ballistic conductance $\sigma_B$ representing the unit conductance times the number of propagating modes is given by: $\sigma_B = {R_B}^{-1} \approx \frac{e^2}{h} \frac{k_F w}{\pi}$, where $k_F$ and $w$ are the Fermi wave vector and the width of the topological insulator channel, respectively. For 3D bulk conduction, $k_F=\sqrt[3]{3\pi^2n}$ where $n$ is the 3D carrier density from all the bulk modes. However, with the chemical potential in the bulk bands, surface conduction may also co-exist along with the bulk conduction, with an associated $k_F=\sqrt{4\pi n_s}$ where $n_s$ is the 2D carrier density of a surface conduction channel.\cite{Culcer2012} Thus, a proper accounting of the distribution of current between bulk and surface conduction channels is actually needed to interpret experimental data when the chemical potential lies in the bulk bands. 

Our studies of ambipolar transport and weak anti-localization at different gate voltages suggest that, near the charge neutrality point, the bulk is depleted and a current flows only through the surface states. In this case, we can calculate $k_F$ using the lowest carrier density near the charge neutrality point ($7.72\times 10^{12}$ $cm^{-2}$ for Dev3 and $4.13\times 10^{12}$ $cm^{-2}$ for Dev4); this corresponds to the residual carrier density from the puddles of electrons and holes created by disorder\cite{Culcer2012,Kim2012}. For the surface state, we determine $k_F =\sqrt{4\pi n_s}=0.070$ \AA $^{-1}$ for Dev3 and $0.051 \AA ^{-1}$ for Dev4. Using $P_{\rm{FM}} \approx 0.51 - 0.63$ for a 20 nm thick Py layer\cite{Vlaminck2008,Haidar2013} and the maximal $\Delta V$ near the charge neutrality point, we estimate the spin polarization of the topological insulator surface state $P_{\rm{TI}} = 0.42 \pm 0.15$ for Dev3 and $0.78 \pm 0.26$ for Dev4. These values are consistent within the error bars and comparable to theoretically calculated values of 0.5\cite{Yazyev2010} or 2/$\pi$\cite{Hong2012}. We note that the estimated numbers are lower bounds since the effective magnetic polarization $P_{\rm{FM}}$ could have lower values for non-ideal \AlO{}/Py junctions. 

\section{Conclusions}

In conclusion, we have carried out electrical measurements of the spin polarization of the surface currents in electrically gated \BST{} devices as a function of the chemical potential. We estimate the spin polarization of the topological surface state to be $0.42 \pm 0.15$ and $0.78 \pm 0.26$ from two devices. Larger spin-dependent voltages were detected as the chemical potential approaches the Dirac point. Our results suggest that to achieve the maximum spin-dependent voltages from 3D topological insulators it is important to avoid the effects from the bulk bands, such as the parallel conduction of the unpolarized bulk carriers or the opposite spin polarization by Rashba spin-orbit interaction. Also, the results point toward a new strategy for electrical control of the magnitude of spin-dependent voltages from the current-induced spin polarization for potential topological spintronics, using gate-tunable topological insulator films.
\bigskip

{\bf Acknowledgement}

We are grateful to Ching-Tzu Chen, Paul Crowell, and Seokmin Hong for valuable comments and discussions. This work was supported by C-SPIN, one of six centers of STARnet, a Semiconductor Research Corporation program, sponsored by MARCO and DARPA. Part of this work was carried out in the College of Science and Engineering Characterization Facility, University of Minnesota, which has received capital equipment funding from the NSF through the UMN MRSEC program under Award Number DMR-1420013. Part of this work was carried out in the College of Science and Engineering Minnesota Nano Center, University of Minnesota and the National Nanofabrication Users Network Facility, Penn State University, both of which receive partial support from NSF through the NNIN program.

%

\clearpage

\begin{table}[t]
	\caption{List of devices with different MTJ areas, MTJ shapes, \BST{} samples, thicknesses of \BST{} film, and thicknesses of \AlO{} tunnel barrier.}
 \begin{tabular*}{0.48\textwidth}{@{\extracolsep{\fill} } c c c c c c }
  \hline \hline
  Device & MTJ Area & MTJ & Sample & \BST{} & \AlO{} \\
	 & ($\mu m^{2}$) & Shape &  & t (nm) & t (nm) \\ \hline
  Dev1 & 50$^2 \pi$ & circle & S1 & 7 & 2.4 \\ 
	Dev2 & 100$^2 \pi$ & circle & S1 & 7 & 2.4 \\ 
	Dev3 & 50$\times$50 & square & S1 & 7 & 2.4 \\ 
	Dev4 & 10$\times$80 & rectangle & S2 & 8 & 2.1 \\ 
	Dev5 & 40$\times$80 & rectangle & S3 & 6 & 2.7 \\ \hline \hline
  \end{tabular*}
\label{Devices}
\end{table}

\clearpage

\end{document}